# IMPROVING THE MANAGEMENT OF COST AND SCOPE IN SOFTWARE PROJECTS USING AGILE PRACTICES

Mariana de Azevedo Santos[1], Paulo Henrique de Souza Bermejo[2], Marcelo Silva de Oliveira[3], Adriano Olímpio Tonelli[2], Enio Júnior Seidel[4]

[1]Mitah Technologies, Lavras, Minas Gerais, Brazil
`mariana@bsi.ufla.br`
[2] Department of Computer Science – Universidade Federal de Lavras (UFLA)
`bermejo@dcc.ufla.br, tonelli@dcc.ufla.br`
[3] Department of Mathematical Sciences – Universidade Federal de Lavras (UFLA)
`marcelo.oliveira@dex.ufla.br`
[4] Universidade Federal do Pampa (UNIPAMPA)
`ejrseidel@hotmail.com`

## ABSTRACT

*While organizations want to develop software products with reduced cost and flexible scope, stories about the applicability of agile practices to improve project development and performance in the software industry are scarce and focused on specific methodologies such as Scrum and XP. Given these facts, this paper aims to investigate, through practitioners' perceptions of value, which agile practices are being used to improve two performance criteria for software projects—cost and scope. Using a multivariate statistical technique known as Exploratory Factor Analysis (EFA), the results suggest that the use of agile practices can be represented in factors which describe different applications in software development process to improve cost and scope. Also, we conclude that some agile practices should be used together in order to get better efficiency on cost and scope in four development aspects: improving (a) team abilities, (b) management of requirements, (c) quality of the code developed, and (d) delivery of software on-budget and on-time.*

## 1. INTRODUCTION

The need for organizations to develop higher quality software products with reduced cost and flexible scope is increasing. This need can be met through the concept of agility, which emphasizes the ability of individuals to respond to changes while de-emphasizes following a detailed plan [1].

Agile practices, consolidated in the 2000s through the creation of the Agile Manifesto [2], have been solidified as a viable alternative for improving the cost and scope of projects in the development cycle of software.

Among the principles advocated in the Agile Manifesto [2] are communication, objectivity, a greater focus on development and customer interaction, conceptual simplicity, high quality, technical excellence, lower costs, dynamism regarding changes to project requirements, flexibility, autonomy, efficiency of development and quick delivery of functional software [3-6]. Interest in these principles is increasing, as organizations and researchers are trying to find solutions to common project problems [7] such as scope and deadline control, code quality, communication efficiency and team cooperation, as well as well as trying to analyze the impact, effectiveness and challenges of software development [5, 8-12]. In fact, for companies, the use of appropriate software development methodology and its practices are one of the most critical issues in recent times [13].

Besides studies involving the application of agile methodologies and their principles, researchers such as Chow and Cao [14], Lee and Xia [15], Abbas et al. [16], So et al. [17], Santos et al. [18], Sletholt et al. [19] and Asnawi et al. [20], have been conducting more thorough studies on some of the general aspects, seeking to understand what kind of attributes are responsible for the success of agile projects. These researchers suggest the existence of factors that promote improvements and successful software development processes. In particular, Chow and Cao [14] describe attributes such as a dynamic environment, proper choice of process management and the use of agile techniques as factors that positively influence the creation of software products.

In order to understand the factors that determine agility dimensions in a project, Lee and Xia [15] suggest a trade-off relationship between response extensiveness and response efficiency of the team. They also suggest that dimensions impact software development agility differently: response efficiency positively affects on-time completion, on-budget completion and the quality of software functionality, while response extensiveness positively affects only the quality of software functionality.

According to Sletholt et al. [19], projects that use agile practices work better with activities related to test and requirement analysis. These activities, if implemented well, can yield good results, achieving affordable cost and flexible scope.

In McHugh et al. [5], the agile teams interviewed stated that using agile practices increased the transparency and visibility of their projects and of the routine activities of their team and their organization. Despite the benefits mentioned, Puhl and Fahney [21] and Eckfeldt [22] highlight the existence of some concerns of customers and developers about the agile approach to projects, concerns such as scope management and have a fixed price contracts. Also, developers are still seeking improvements regarding scope, cost estimate, and overall project performance while trying to promote more positive and productive relationships with customers and to deliver better products [21, 22].
Besides the benefits and challenges reported in these studies about agile development, the focus of these analysis were in agile principles and specific methodologies, such as XP and Scrum, and not on the use of agile practices in general as applied nowadays [3].

Moreover, studies by Abbas et al. [16], So et al. [17] and Asnawi et al. [20] go beyond the ones mentioned above, as they used the Factor Analysis technique to identify the effectiveness of agile practices and their social-psychological effects.

The results of the study of So et al. [17] suggest eight factors extracted to identify the social-psychological e ects in eight agile practices commonly used in the IT field. However, the practices used in the Factor Analysis were not generalized and the extracted factors represent the same eight practices proposed in the study.

On the other hand, the results of Abbas et al. [16] propose 15 factors, originating through practices evaluated in a survey, which explain contributions in several IT areas such as process/governance, quality assurance, iterative and incremental development and team communication but not directed at project performance aspects such as cost, quality, deadlines and scope.

Asnawi et al. [20], examined 27 variables of agile practice in Malaysian companies, applying Factor Analysis, and found eight factors. These extracted factors explain the practitioners' negative perceptions about the importance of agile development technical aspects in adoption process, valuing more human aspects such as customer satisfaction and collaboration among developers and customers. The research had a sample of 88 complete evaluations.

In another study, Asnawi et al. [23] investigated, through qualitative research, Malaysian adopters' perceptions of agile methods. The researchers founded that people in management roles have difficulties accepting agile methods and, in general, need to see something working and to hear success stories from agile users. This study, as with Asnawi et al. [20], mentioned above, was limited to Malaysia.

Filling the need for deeper analysis of agile development practices in projects, Santos et al. [18] analyzed practitioners' perceptions of the impact of agile practices, using factor analysis and considering different methodologies such as Feature-driven development (FDD), Extreme programming (XP), Scrum, Crystal Methodologies, Dynamic systems development method (DSDM), Test-driven development (TDD). With a sample of 109 complete evaluations, Santos et al [18] found six factors that should be considered as a successful attribute for achieving higher quality in software products.

Despite the Santos et al. [18] results, success attributes in software development can be described not only by quality but also by scope and cost [15, 25]. With this in mind, it is important to note perceptions of other crucial aspects of the project life cycle, such as development cost, or the time and effort used to estimate and implement required tasks [15], and scope, or the main parameters to be controlled on a project [16].

Our goal was to analyze which agile practices are used, according to practitioners' perceptions, to improve two performance criteria in software projects, cost and scope, and focuses on the following research question:

(a) Which of the main agile practices can contribute to better efficiency on cost and scope in software development?

With this question in mind, our paper is organized as follows: Section 2 presents our research methodology, including a description of the methods used to analyze data. Section 3 reports our findings and is divided into three parts - data validation, or how the data were validated by statistical techniques; an explanation of the cost factors; and an explanation of the scope factors. Section 4 discusses the factors extracted and the impact of these on each criterion. Section 5 presents our conclusions. And Section 6 acknowledges research limitations and provides recommendations for future works.

## 2. METHODOLOGY

The research methodology is quantitative and descriptive, focusing on the identification and analysis of factors and variables related to the phenomenon or process [26].

Our research process was developed in three stages: (1) a literature review where we identified the agile practices used most and classified them into three groups: management practices, development practices (rules, coding, testing and integration), and planning and monitoring practices [18]; (2) data pre-processing (see Section 2.1); and (3) exploratory factor analysis (see Section 2.2).

The data were collected over four months, from users of social network communities and discussion lists about software engineering, via the online tool *SurveyMonkey*, from users in Brazil and throughout the world.

The survey had 17 questions distributed into four groups, as Table 1 shows: characterization of the participating company and individual (demographic and skills information), agile practices usage analysis, part 1 (information about adoption and personal agile practices knowledge), agile practices usage analysis, part 2 (evaluation of practices on a 6-point Likert scale based on the cost and scope) and agile practices usage analysis, part 3 (benefits and challenges of agile adoption).

Table 1 Survey roadmap

| Questionnaire items groups | Number of items | Types of variables evaluated in the items | Likert scale? | Example of item |
|---|---|---|---|---|
| 1. Presentation of the company and the participant | 6 | Nominal and ordinal categorical | No for all items | 3. Role in the company (Select the role that best fits the activities conducted by you in the company): |
| 2. Agile practices usage analysis - part 1. | 3 | Ordinal categorical | No for all items | 7. How long the company adopts agile methods? |
| 3. Agile practices usage analysis - part 2. | 5 | Ordinal categorical | Yes for all items | 12. Which agile practices related to organization of teams and feedback are used?… What is the perceived value of this practice in relation to project cost? |
| 4. Agile practices usage analysis - part 3. | 3 | Nominal and ordinal categorical | Yes for some of the items (and no for the others) | 17. Which of the following do you consider a potential benefit in adopting agile methods? |
| **Total** | **17** | | | |

We obtained 109 responses from around the world, as represented in Figure 1, with 77 respondents from Brazil, 19 from the United States, and one respondent each from Argentina, Australia, Belgium, Colombia, Denmark, France, Germany, India, Italy, Poland, Switzerland, Serbia, and Venezuela.

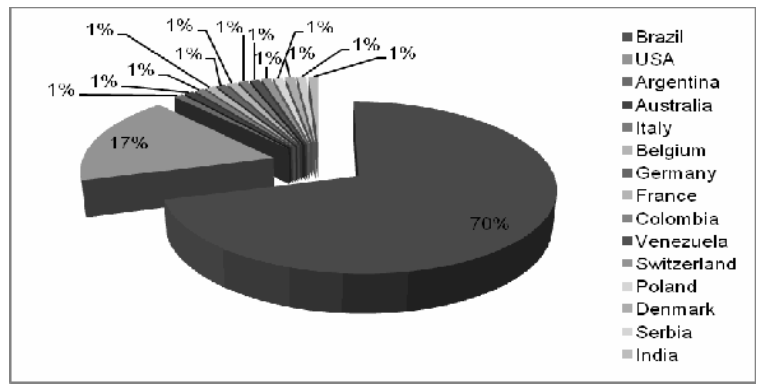

Figure 1 Countries that participated in the survey

The participants were predominantly project managers (27.5%), technical leaders (11%), and project leaders (10.1%). None were customers. Responses about the perceived value of the use of agile practices were organized into three groups, representing the main stages of the project life cycle: (1) project management practices, (2) development practices and (3) planning and tasks monitoring practices.

The practices were selected from a literature review of agile methodologies and were the most used in the software industry, as confirmed by the survey's first part represented in Figure 2.

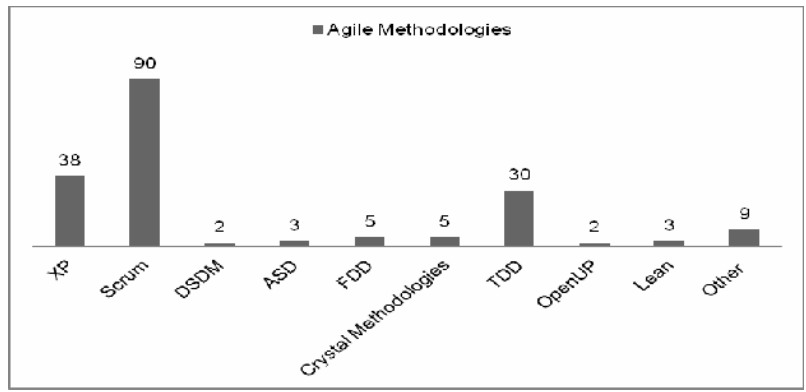

Figure 2 Agile methodologies most used according to questionnaire results

In the survey's second part, participants indicated how much they valued an item following a Likert scale of six points: very high (6), high (5), and satisfying (4) to evaluate practices that positively impacted their experience, and regular (3), low (2), and very low (1) to evaluate practices that negatively impacted their experience. Each item evaluated on the Likert scale (numbered by i = 1, 2, 3 ...) received a score $X_i$. Because of the adoption of a non-deterministic design sampling (observational), $X_i$ was considered a random variable with possible values in the set of {1, 2, 3, 4, 5, 6}.

Since this was an assessment focused on tacit, not explicit, knowledge the participants were not obligated to evaluate all practices, but only those that they had used in their work environments. This measure was taken to obtain answers not biased by certain circumstances, such as a lack of experience or knowledge of a specific practice.

From the point of view of software engineering, this type of research, using a value perspective, provides a good way to analyze the product development process and the value of the practices applied. It also identifies the best alternatives for software companies for creating business strategies and solutions that achieve long-term profitable growth and sustainable competitive advantages [27].

Data analysis was performed in two parts, using two different perspectives. The first part, pre-processing, used descriptive data analysis while the second part, exploratory analysis, used multivariate data analysis.

## 2.1. First part: data pre-processing

During the first phase, descriptive analysis techniques were used to perform a data pre-processing. The objective was to normalize the data and construct a formal database.

Thus, the first step was to run a treatment of missing data. As the practitioners' were not obligated to answer all of the questions in the survey, in other words, the missing values in each variable are independent in the cases and occur randomly, without forecasts, the sample was identified as missing completely at random (MCAR) and for these cases, the treatment was to calculate the mean for each question, substituting in each case the missing value [28, 29].

The second step was to eliminate the outliers, which were the observations showing a substantial discrepancy toward others or were inconsistent or extreme compared to the other results [29]. In this case, we eliminated extremely positive ("6 - very high") or negative ("1 - very low") evaluations on a particular practice that disagreed when compared with the other cases.

Finally, the third step was to perform descriptive diagnostics on the sample, such as extracting demographic and profile information, and verify if the sample size is valid for using multivariate techniques.

## 2.2. Second part: Exploratory factor analysis (EFA)

Following the pre-processing phase, we used a multivariate technique named Exploratory Factor Analysis (EFA) to exploit the results obtained by the variables $X_i$ evaluated on a Likert scale and to propose solutions to the research problem: which of the main agile practices can contribute to better efficiency on cost and scope in software development.

In theory, EFA is a technique for exploratory data analysis consisting in data reduction or a structure simplification, to describe, if possible, the ratio of the covariance among the many variables in random and unobserved quantities named factors [28].

The use of this technique was motivated by the argument that variables can be grouped based in the correlation degree, in other words, the strength and direction of the relationship among variables.

In this study, we are going to test whether a correlation exists between the performance criteria for each agile project, such as cost and scope, and the use of some agile practices. It means that in each of these criteria, there are factors that explain the use of certain practices together, according to the practitioners' perception.

In particular, the EFA model is represented by equations (1):

$$X_1 = {}_{11}F_1 + {}_{12}F_2 + \ldots + {}_{1m}F_m + {}_1$$
$$X_2 = {}_{21}F_1 + {}_{22}F_2 + \ldots + {}_{2m}F_m + {}_2 \qquad\qquad (1)$$
$$\vdots$$
$$X_p = {}_{p1}F_1 + {}_{p2}F_2 + \ldots + {}_{pm}F_m + {}_p$$

In this model, the coefficient $_{ij}$ is named the (factor) loading of the *i*th variable in the *j*th factor, where the letters *i* and *j* are integer index 1, 2, 3 ... , and $L_{(p \times m)}$ is the factor matrix with loadings. In this context, the factor analysis model assumes that these variables show a linear relationship with the new random variables $F_n$, where n = 1, 2, 3 ... m. The vector $_{(px1)}$ represents the random errors associated with measurements [23].

This initial factor matrix, which indicates the relationship among the variables studied, rarely results in factors that can be interpreted [29].

However, an analysis becomes feasible and even useful due to its ability to produce interpretable factors through methods of matrix rotation which transform the matrix of factors into a rotation matrix that is maximized, meaningful, simpler and easier to interpret [30, 31].

We used an orthogonal rotation method named Varimax, which is the most widely used rotation method, by maximizing the sum of variances of required loading of a factor on all variables in the factor matrix, that is, how the variables measure the same concept [28, 32].

The factors found were defined by the value of their index, which was dependent on sample size. For this research, the practices that had factor loadings of 0.55 or higher were considered significant because of the sample size of 109 respondents [29]. The validation tests were made using the Kaiser-Meyer-Olkin (KMO) method and latent root calculation [28, 29].

# 3. RESULTS

## 3.1. Exploratory factor analysis reliability

The agile practices used to conduct the survey were those listed by our literature review to practitioners' indicates those that, best correspond to the reality of the current software industry.

The data collected are shown in Figure 3. The objective of this step was to obtain a list of the main practices used in order to make the sample more significant for the second stage, the EFA itself. Thus, we considered only the practices that had a number of evaluations (n ≥ 35). This criterion was considered best because the analysis required a significant sample to work.

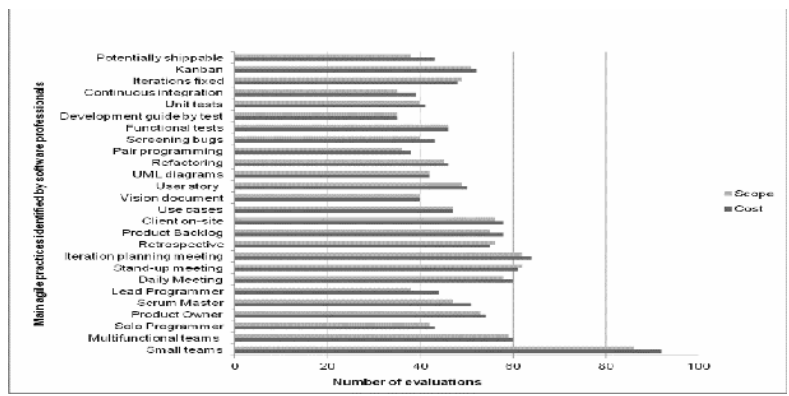

Figure 3 Group of practices most valued based on this research

Once the main practices used in the software industry were identified, the data were ready to be tested by the EFA.

The data validation test was generated using Bartlett's sphericity test, to measure the presence of correlations between the variables $X_i$ and the KMO method, a test of sampling adequacy, the objective of which was to quantify the degree of inter-correlation among the variables and the appropriateness of factor analysis [29]. For each criterion evaluated, a KMO index was found and, according to Maroco [32], each value between 0.5 (reasonable) and 0.9 (excellent) corresponded to a positive recommendation to proceed with an analysis of the sample.

Table 2 KMO Coefficient of Factorial Analysis and Bartlett's sphericity test for representative sampling of cost and scope criteria

| Validity tests used | Results | |
|---|---|---|
| | Cost | Scope |
| Measure of sampling adequacy *Kaiser-Meyer-Olkin* | 0.870 | 0.760 |
| *Bartlett's* sphericity test *($Q^2$ Approximation)* | 26.494 | 7.858 |
| *df* * | 300 | 300 |
| *sig.* ** | 0.00 | 0.00 |

\* Degrees of Freedom
\*\* Significance

In the validity tests, described in Table 2, indicated that for cost criterion, with index $KMO_C = 0.870$, approximately 0.9, validated the analysis with a great recommendation. For the scope test, with index $KMO_S = 0.760$, approximately 0.8, the information validated the analysis with a good recommendation. The Bartlett's test, in both cases, showed a significance value of p <0.0001, and therefore we concluded that the variables $X_i$ were significantly correlated [29, 32].

## 3.2. Factors Extraction

There are several methods described in the literature for extracting factors in the data. Our choice of method was based on two criteria: (1) the factor analysis objective and (2) previous knowledge about the variance on variables [29, 33].

The objective of the use of factor analysis in this research was to explore the data to generate future hypotheses about cost and scope improvements using agile practices. In this case, we chose the extraction method Principal Component Analysis (PCA) because this method explains the total data variance represented in the variables in data reduction to factors [28, 29].

This definition means that the PCA transforms the data to a new coordinate system ensuring the highest variance for any projection data for variables, which define the factors, in descending order. It is also the model most used in Factor Analysis, especially in statistical packages such as Statistical Package for the Social Sciences (SPSS) [29, 33].

Regarding the PCA results for cost criterion, six factors were extracted explaining 64.951% of this data variance. For the scope criterion, seven factors were extracted, explaining 60.893% of this data variance. Beyond the variance as an acceptable parameter for the analysis, it was also necessary to consider the latent root criterion, where eigenvalues should be greater than or equal to 1. In both cases this result was confirmed [28, 29, 32].

With the factors discovery, the solution of each analysis was described in the following subsections. Each Factor Analysis had agile practices of higher perceived value in cost and scope criteria, selected by the higher loadings, as shown in Table 3 and Table 4.

### 3.3. Factors in cost criterion

With the evaluation of practitioners' perceptions of value of the cost criterion, it was possible to extract six factors. Each extracted factor received a set of practices, described in Table 3, which had a higher correlation, defined by the loadings, and was used more frequently to get solutions in software projects related to improvements in cost.

Table 3 Factors loading of a representative sample of agile practices under the cost criterion (bold numbers are chosen as relevant)

| Agile practices | F1 | F2 | F3 | F4 | F5 | F6 |
|---|---|---|---|---|---|---|
| Small teams | 0.156 | **0.594** | 0.049 | **0.560** | -0.078 | 0.021 |
| Multifunctional teams | 0.222 | **0.661** | 0.137 | 0.483 | 0.022 | 0.066 |
| Solo programmer | -0.037 | -0.003 | -0.027 | 0.044 | 0.005 | **0.815** |
| Product owner | 0.079 | **0.737** | 0.301 | 0.030 | 0.209 | 0.221 |
| Scrum master | 0.486 | **0.640** | 0.012 | 0.070 | 0.168 | 0.154 |
| Lead-programmer | 0.254 | **0.564** | 0.224 | 0.090 | -0.030 | 0.121 |
| Daily meeting | **0.747** | 0.131 | 0.068 | 0.277 | 0.103 | 0.050 |
| Stand-up meeting | **0.721** | 0.154 | 0.213 | -0.007 | 0.025 | 0.074 |
| Iteration planning meeting | **0.726** | 0.239 | 0.259 | 0.171 | 0.051 | 0.079 |
| Retrospective | **0.726** | 0.222 | 0.276 | 0.230 | -0.036 | 0.109 |
| Product backlog | **0.557** | 0.438 | 0.163 | 0.129 | 0.262 | 0.101 |
| Client on-site | 0.160 | **0.587** | 0.176 | 0.272 | 0.151 | 0.263 |
| Use cases | 0.078 | 0.022 | 0.062 | 0.024 | **0.761** | 0.366 |
| Vision document | 0.508 | 0.036 | 0.139 | 0.003 | 0.390 | 0.363 |
| User story | **0.584** | 0.365 | 0.190 | 0.272 | 0.173 | 0.288 |
| UML diagrams | 0.022 | 0.159 | -0.119 | 0.132 | **0.683** | 0.268 |
| Refactoring | 0.173 | 0.125 | **0.628** | 0.453 | -0.061 | 0.149 |
| Pair programming | 0.232 | 0.137 | 0.541 | 0.187 | 0.453 | 0.042 |
| Screening bugs | 0.147 | 0.054 | 0.210 | **0.607** | 0.234 | 0.262 |
| Functional tests | 0.264 | **0.558** | 0.195 | -0.004 | 0.460 | 0.197 |
| Development guide by test | 0.372 | 0.300 | **0.693** | 0.125 | -0.062 | 0.011 |
| Unit tests | 0.269 | 0.497 | **0.564** | -0.152 | 0.038 | 0.291 |
| Continuous integration | 0.400 | 0.239 | **0.635** | 0.140 | 0.086 | 0.124 |
| Iterations fixed | **0.579** | 0.161 | 0.424 | 0.130 | 0.106 | 0.145 |
| Kanban | 0.406 | 0.193 | 0.090 | **0.619** | 0.085 | 0.150 |
| Potentially shippable | 0.157 | 0.191 | 0.357 | 0.450 | 0.429 | 0.045 |

The statistical interpretation of the results displayed in Table 3 above originated from the EFA model (equations (1)):

$$X_1 = 0.156F_1 + 0.594F_2 + 0.049F_3 + 0.560F_4 - 0.078F_5 + 0.021F_6 + \epsilon_1$$
$$X_2 = 0.222F_1 + 0.661F_2 + 0.137F_3 + 0.483F_4 + 0.022F_5 + 0.066F_6 + \epsilon_2$$
$$X_3 = -0.037F_1 + (-0.003)F_2 + (-0.027)F_3 + 0.044F_4 + 0.005F_5 + 0.815F_6 + \epsilon_3$$
$$X_4 = 0.079F_1 + 0.737F_2 + 0.301F_3 + 0.030F_4 + 0.209F_5 + 0.221F_6 + \epsilon_4$$
$$\vdots$$
$$X_{26} = 0.157F_1 + 0.191F_2 + 0.357F_3 + 0.450F_4 + 0.429F_5 + 0.045F_6 + \epsilon_{26}$$

The variable symbols could be interpreted as:

$X_1$: practitioners' perceptions about the impact of the use of agile practice "Small teams" under the cost criterion.

$X_2$: practitioners' perceptions about the impact of use of agile practice "Multifunctional teams" under the cost criterion.

$X_3$: practitioners' perceptions about the impact of use of agile practice "Solo programmer" under the cost criterion.

$X_4$: practitioners' perceptions about the impact of use of agile practice "Product owner" under the cost criterion.

...

$X_{26}$: practitioners' perceptions about the impact of the use of agile practice "Potentially shippable" under the cost criterion.

The factors presented below were identified through the factor analysis and represent six topics related to cost in software project management based on software agile practices. They are:

**Factor 1 - Team's interaction:** this first factor represents high factorial weights in agile practices related to the agile meetings in the planning and development stage, using the following practices: planning meetings, daily scrum, stand-up meetings, retrospectives, user stories and fixed iterations. In this process, the team defines which story (user story) should be implemented and completes the task estimation during the Sprint planning meeting or iterative cycle planning meeting. The team's progress on tasks is monitored through daily and stand-up meetings. At the end of the Sprint, a meeting named retrospective is held, during which the cycle's strengths and weaknesses are identified and each team member suggests improvements. According to the practitioners' perceptions, meetings are important to the project's stages not only to add value to the teams' work, but also to help identify possible obstacles and avoid problems arising from lack of communication and from the task estimation errors, reducing implementation costs and rework.

**Factor 2 – Customer's on-site on features test:** this second factor had high factorial weights in agile practices related to agile teams' hard involvement on functional test techniques, using the practices: multi-functional teams, led by a scrum master, with the customer's on-site participation in the features test, giving needed feedback. In this factor, practitioners' perceptions indicate a lack of distance between the development team, management and the customer, making information more decentralized and creating a strong culture of knowledge management among members.

**Factor 3 - Clean code development:** this third factor had high factorial weights in agile practices related to the development stage. For this factor, development begins with test cases. As the code is developed, unit tests are performed and primary errors are corrected, followed by refactoring if necessary. This code, tested and validated by the team, is integrated into the project scope through a continuous integration practice, i.e., each feature validated will be integrated into the code by the developers. Under this scenario, cost is optimized from the beginning of the project, preventing serious flaws and allowing the project to be adjusted according to a customer's changing requirements.

**Factor 4 – Time response:** this fourth factor had high factorial weights in agile practices related to the team's quickness in response to changing requirements. In this factor, the team monitors your progress through the stories or simple tasks come from the product backlog as new requirements identified through the acceptance test into a scrum task board. For practitioners, it is understood that the faster the team can adapt to these changes, the lower the cost of the project schedule.

**Factor 5 - Consistent domain model:** this fifth factor had high factorial weights for practices related to creation of the project domain model. The most valued practices to the users in this factor are the use of case diagrams and class diagrams. A domain well modelled with low coupling and dependency, as a planning meeting's product, may represent fewer project cost risks.

**Factor 6 - Solo development:** this sixth factor represents high factor weights in solo programmer practice. This factor shows solo developer role as the only one responsible for assignments and tests. This discrepant factor should be a greater convenience to some practitioners when dealing with task estimations in short-term project, once the developer knows your abilities, with no impediments related to dependence on other team members.

### 3.4. Factors in scope criterion

After the evaluation of agile practices under the cost criterion, participants were invited to evaluate agile practices under the scope criterion. As a result of this evaluation, seven factors were extracted. For each factor extracted a set of practices, as described in Table 4, shows a correlation, defined by the loadings, and are used more frequently to optimize the project scope.

Table 4 Factors loading of a representative sample of agile practices under the scope criterion (bold numbers are chosen as relevant)

| Agile Practices | F1 | F2 | F3 | F4 | F5 | F6 | F7 |
|---|---|---|---|---|---|---|---|
| Small teams | 0.121 | 0.035 | **0.749** | 0.096 | 0.087 | 0.097 | 0.012 |
| Multifunctional teams | 0.529 | 0.037 | 0.302 | 0.099 | 0.226 | 0.350 | 0.058 |
| Solo programmer | 0.086 | 0.144 | 0.075 | -0.172 | -0.016 | 0.066 | 0.135 |
| Product owner | 0.148 | 0.119 | 0.439 | 0.508 | -0.029 | 0.002 | 0.204 |
| Scrum master | 0.137 | 0.259 | **0.582** | 0.137 | 0.171 | 0.224 | 0.483 |
| Lead-programmer | 0.149 | -0.056 | 0.092 | 0.028 | 0.136 | 0.163 | **0.833** |
| Daily meeting | 0.127 | 0.087 | -0.072 | -0.031 | 0.528 | 0.175 | 0.270 |
| Stand-up meeting | 0.176 | 0.077 | 0.153 | -0.042 | **0.786** | 0.041 | 0.016 |
| Iteration planning meeting | 0.068 | -0.042 | 0.375 | 0.307 | 0.438 | 0.280 | 0.219 |
| Retrospective | 0.199 | 0.306 | 0.133 | 0.396 | 0.422 | 0.150 | 0.228 |
| Product backlog | 0.058 | 0.102 | 0.132 | **0.761** | 0.093 | 0.072 | 0.176 |
| Client on-site | 0.113 | 0.023 | 0.068 | **0.605** | -0.071 | 0.256 | 0.288 |
| Use cases | 0.053 | **0.752** | -0.100 | 0.082 | 0.142 | 0.023 | 0.125 |
| Vision document | -0.060 | **0.769** | 0.023 | 0.251 | 0.143 | 0.031 | 0.050 |
| User story | 0.046 | 0.076 | 0.540 | 0.273 | 0.153 | 0.234 | 0.011 |
| UML diagrams | 0.097 | **0.673** | 0.211 | -0.174 | -0.001 | 0.142 | 0.172 |
| Refactoring | **0.555** | 0.368 | 0.210 | 0.067 | 0.160 | 0.143 | 0.183 |
| Pair programming | 0.270 | 0.163 | 0.042 | 0.217 | 0.234 | 0.085 | 0.304 |
| Screening bugs | 0.300 | **0.613** | 0.222 | 0.030 | -0.283 | 0.048 | 0.288 |
| Functional tests | **0.557** | 0.532 | 0.138 | 0.081 | -0.037 | 0.311 | 0.175 |
| Development guide by test | **0.871** | -0.006 | -0.027 | 0.094 | 0.127 | 0.011 | 0.186 |
| Unit tests | **0.882** | 0.119 | 0.090 | 0.043 | 0.127 | 0.017 | 0.053 |
| Continuous integration | 0.401 | -0.020 | 0.158 | 0.273 | 0.434 | 0.281 | 0.022 |
| Iterations fixed | 0.235 | 0.343 | -0.130 | 0.242 | 0.163 | 0.505 | 0.264 |
| Kanban | 0.238 | 0.139 | 0.376 | -0.234 | -0.076 | 0.504 | 0.114 |
| Potentially shippable | 0.043 | 0.049 | 0.101 | 0.156 | 0.138 | **0.765** | 0.044 |

The statistical interpretation of the results displayed in Table 4 above originated from the EFA model (equations (1)), with similar interpretation of Table 3:

$$X_1 = 0.121F_1 + 0.035F_2 + 0.749F_3 + 0.096F_4 - 0.087F_5 + 0.097F_6 + 0.012F_7 + \quad _1$$
$$X_2 = 0.529F_1 + 0.037F_2 + 0.302F_3 + 0.099F_4 + 0.226F_5 + 0.350F_6 + 0.058F_7 + \quad _2$$
$$X_3 = 0.086F_1 + 0.144F_2 + 0.075F_3 + (-0.172)F_4 + (-0.016)F_5 + 0.066F_6 + 0.135F_7 + \quad _3$$
$$X_4 = 0.148F_1 + 0.119F_2 + 0.439F_3 + 0.508F_4 + (-0.029)F_5 + 0.002F_6 + 0.204F_7 + \quad _4$$
$$.$$
$$:$$
$$X_{26} = 0.043F_1 + 0.049F_2 + 0.101F_3 + 0.156F_4 + 0.138F_5 + 0.765F_6 + 0.044F_7 + \quad _{26}$$

The variable symbols could be interpreted as:

$X_1$: practitioners' perceptions about the impact of the use of agile practice "Small teams" under the scope criterion.

$X_2$: practitioners' perceptions about the impact of the use of agile practice "Multifunctional teams" under the scope criterion.

$X_3$: practitioners' perceptions about the impact of the use of agile practice "Solo programmer" under the scope criterion.

$X_4$: practitioners' perceptions about the impact of the use of agile practice "Product owner" under the scope criterion.

...

$X_{26}$: practitioners' perceptions about the impact of the use of agile practice "Potentially shippable" under the scope criterion.

The factors presented below were identified through the factor analysis and represent seven topics related to scope in software project management based on software agile practices. They are:

**Factor 1 - Development guide by test:** this first factor had high ratings in the following practices: developing test cases, unit tests and refactoring. In this factor, the developers creates tests for features that isn't complete yet, writes code and testing this new features until it becomes errorless and functional, improving the code quality through the refactoring practice. This perception denotes a high value to the project insofar as it avoids delays in the planned scope originating from poor code maintenance, which in turn helps to improve code quality and to get maturity on tests.

**Factor 2 - Objective documentation:** the second factor represents the use of the following practices: vision document, Unified Modeling Language (UML) diagrams, screening errors and use cases. The context described in this factor includes the use of UML diagrams to represent the classes and use cases of the system, and modifying them frequently according to the feedback of acceptance tests made by the customer and registered in the vision document. The vision document is streamlined and simplified documentation written in common language according to the customer's needs and what the software functionalities should be. This factor contributes positively to the scope criterion, and provides consistency and flexibility according to practitioners' perceptions.

**Factor 3 - Small teams led by the facilitator:** this third factor had high factorial weights in practices related to the formation of small teams led by the team's facilitator. The facilitator works to keep the team focused on the Sprint scope, helping to solve problems on the tasks and giving recommendations on best practices in tasks development. In case tasks had problems in Sprint, the facilitator allowed postponing it in Sprint backlog. This role functions as a mediator of the team's problems.

**Factor 4 - Features defined by the customer:** this fourth factor represents the practitioner's appreciation for the customer by joining the team for the backlog development activities, activities such as choosing and prioritizing tasks which will add greater value to the product. The customer's on-site behaviour contributes, as with the third factor mentioned above, to the team members remaining focused on the scope, responding better and negotiating deadlines in case of changing requirements.

**Factor 5 - Stand-up meetings:** the fifth factor shows positive practitioners' perceptions of the stand-up meetings practice in an attempt to improve the flow of communication among team members and to update the team's knowledge of the Sprint's progress. For the scope criterion, this factor contributes to a more informal and frequent status presentation of the project. It allows members to understand others' responsibilities and what tasks are done or not, without consulting the project schedule document with unnecessary frequency.

**Factor 6 - Frequent releases:** the sixth factor shows the practitioners' perceptions in the practice product potentially shippable, known as a functional release. The practitioners' believe that developing a functional release at the end of each Sprint and presenting it to the customer can contribute to a planned scope because the software must be delivered in functional parts. This factor also decreases the risk of errors and of having features poorly implemented throughout the entire project.

**Factor 7 - Lead-programmer:** the seventh factor had the highest factorial weights fin the practice: lead-programmer role on team. This factor explains practitioners' positive perceptions of the value of this role, which involves working as a developer but also being responsible for helping the project manager develop consistent and clear software architecture and solve problems. In the scope criterion, the role of the lead-programmer presents a solution to improve features development and helps to define it better. In addition, this role also involves being a leader of the development team and providing technical support to team members when necessary.

# 4. DISCUSSION

From the results of the Exploratory Factor Analysis (EFA), we can describe two small scenarios where certain practices are used based on the perceptions and experiences of research participants. Thus, when these practices are grouped into factors, in order to organizations optimize your process, based on the good results in a given criterion. Describing the use of practices in these factors, we find that in general, practitioner's' perceptions about using agile practices to reduce cost and flexible scope can be summarized in the following in four development aspects: improving (a) team abilities, (b) requirements management, (c) quality of the code developed, and (d) delivering of software on-budget and on-time.

Regarding the first development aspect, three factors were observed in cost criterion. The factors team's interaction, customer's on-site on features test and lead-programmer development have a positive impact on cost, improving the team's leadership abilities and creating high-performance teams with knowledge enough to work in several aspects on Sprint's project. By taking these measures, high costs coming from poor coverage tests or lack of knowledge by the team may be drastically reduced.

For the scope criterion, multifunctional teams led by a facilitator reinforce the presence of a leadership representative on teams, helping in issues as customer requirements understanding, assisting in the scope's project definition. The facilitator usually has more experience than team members and can have a better understanding of selected tasks, and can define what should be prioritized and implemented before the deadline, with a more realistic understanding of time and budget.

In relation to the team's communication, we found factors that address the communication of two strands: time response and team meetings to recognize the scope implemented. The first factor, time response, is how skilled the team is in communication and self-organization to improve features with errors, whether those selected in the task board or arising out of acceptance tests. Teams that apply practices of the time response factor tend to reduce cost modifications because they have excellent efficiency, with modifications producing limited impact on the project.

The second factor, stand-up meetings, relates to how skilled the team is in: (1) communicating and organizing the implementation of the project scope distributed in the Sprints, (2) determining in the meetings who will implement tasks, and (3) resolving problems and difficulties. The use of these practices contributes to a better knowledge of what should be implemented.

The development aspect (b), requirements management, it is defined by one factor in the cost criterion: consistent domain model. A domain model can be defined as a model responsible for defining the behaviour and data of a project. It is argued that the scope represented in UML diagrams and detailed in the planning meeting are key factors contributing to the success of the project regarding development costs. Working within a flexible scope, the domain modelling will decrease the total cost of the changes required by the customer during the build of the software product.

The development aspect (c), the quality of the code developed, is affected by three factors: (1) clean code development, (2) development by test cases, and (3) objective documentation. In a comprehensive way, all these factors approach that in agile development, the company develops in small portions with integrated functional code named release. Another point that should be observed is that between the criteria analyzed, development based on test cases is present. This can mean developing written tests to locate failures in the code and to fix them, not only impacting the quality of the code but also the cost of implementation, the delivery deadline control and the scope control.

The last development aspect, (d) delivery of software on-budget and on-time, has a similar behaviour in both criteria studied. The common point among the described factors is that releases with a pre-set time, defined by effort estimation practices and frequent deliveries, assist in forming a more cohesive product, a product that is both high quality and functional. Therefore, with the fixed-cycle time, the team is attentive to the term and knows better what it should deliver and what it can deliver for the Sprint, avoiding mistakes and generating knowledge for the next release. The factor solo development in this aspect suggests that respondents had a positive experience with the practice in projects with short deadlines.

These findings should be interpreted as recommendations for practitioners, meaning that it is possible to get better and satisfactory results in cost and scope using these practices to improve team management abilities, to organize management of requirements, to improve code quality and to deliver software on budget and on-time.

A team with high performance should be functional, write good and maintainable code, be creative and innovative enough to better understand their customer's needs and prioritize tasks in the project scope that will add value. These abilities are essential for a team to work better in projects with a flexible scope.
Moreover, costly projects usually have problems with poor code and misinterpreted requirements which often produce bugs, bugs needing to be fixed in numerous Sprints that end up not being productive and negatively influencing software delivery.

By correlating the results of this paper with previous works, we have successfully achieved our goal through the exploratory factor analysis, discovering factors that explain practitioners' perceptions of the main agile practices responsible for obtaining results efficiency on cost and scope in the current software industry.

## 5. CONCLUSION

With the factor analysis study, we were able to identify small groups based on practitioners' perceptions and experiences, showing how companies and individuals are applying agile practices in their environment. Thus, represents how these practices can be grouped into structures named factors making possible to organizations' improving teams' abilities to plan scopes successfully, to develop costs sustainably, and to even optimize defined and managed processes.

This article's goal was to analyze the agile practices used to improve two performance criteria in software projects, seeking to answer the following research question: which of the main agile practices can contribute to better efficiency on cost and scope in software development?

Regarding the question, the six factors extracted for cost criterion and seven factors extracted for scope criterion attained by Exploratory Factor Analysis, suggest factors derived from practitioners' perceptions of value, describing which agile practices could be clustered or using in the same context in order to get better efficiency on cost and scope in four different aspects: improving (a) team abilities, (b) management of requirements, (c) quality of the code developed, and (d) delivery of software on-budget and on-time. These aspects summarize practitioners' perceptions about using agile practices in software projects to get reduced cost and flexible scope.

Another important conclusion was that results can be more widespread by analyzing a set of agile practices rather than a specific methodology. This measure could represent a better path to achieving mature results on projects using the agile development approach.

Derived from an acceptable, worldwide statistical sample, these results should be interpreted as recommendations for success in terms of cost and scope and are applicable to projects all around the world.
Some practices, however, could not be assessed because of the low amount of responses, somewhat reducing the scope range of practices used in the research. The low amount of responses can be explained by the number of respondents or by a lack of knowledge among respondents about the use of certain practices.

In summary, we expect that the study results encourages further quantitative research in software engineering field, so that organizations and the academic community can improve their investments, resources and efforts in software development using the agile approach to implement creative solutions to get affordable cost and flexible scope.

## 6. LIMITATIONS AND FUTURE WORK

This article has limitations concerning the interpretations of results. First, the sample collected is limited, implying limited inferences, and so the results should be interpreted as *initial* recommendations regarding agile software development. Future work is suggested, work involving more in-depth statistical research and analysis, using confirmatory factor analysis for example. In addition, we also suggest conducting qualitative studies to further investigate these factors, including an investigation of what causes and effects of these factors through a case study.

## ACKNOWLEDGEMENTS

The authors thank The Minas Gerais State Research Foundation (FAPEMIG), The Coordination for the Improvement of Higher Level or Education (CAPES) and The National Council of Technological and Scientific Development (CNPq) for their financial support.

## REFERENCES

[1] Karimi, Z, Behzady, S, Broumandnia, A (2012) "Achieving the Benefits of Agility in Software Architecture-XP". International Journal of Computer Science & Information Technology (IJCSIT), Vol 4, No 5, October, pp.103-114.

[2] Beck, K., Beedle, M., Bennekum, A. van, Cockburn, A., Cunningham, W., Fowler, M., Grenning, J., Highsmith, J., Hunt, A., Jeffries, R., Kern, J., Marick, B., Martin, R. C., Mallor, S., Schwaber, K., Sutherland, J. (2001) "Agile Manifesto". http://www.agilemanifesto.org/. Accessed in 18 August 2012.

[3] Cao, L, Ramesh, B (2007). "Agile Software Development: Ad Hoc Practices or Sound Principles?" IT Professional 9, 2, pp. 41-47.

[4] Dybå, T, Dingsøyr, T (2008) "Empirical studies of agile software development: A systematic review". Informatics, Software Technology. 50, 9-10, pp. 833-859.

[5] Sidky, A, Arthur, J, Bohner, S (2007) "A disciplined approach to adopting agile practices: the agile adoption framework". Innovations in Systems and Software Engineering, Springer London, Volume 3, Number 3, 203-216. doi: 10.1007/s11334-007-0026-z

[6] McHugh, O, Conboy, K, Lang, M (2012) "Agile Practices: The Impact on Trust in Software Project Teams," Software, IEEE, vol.29, no.3, pp.71-76, May-June 2012, doi: 10.1109/MS.2011.118

[7] Salo, O, Abrahamsson, P (2008) "Agile methods in European embedded software development organizations: a survey study of Extreme Programming and Scrum". Published in IET Software, 2, (1), pp. 58-64.

[8] Mann, C, Maurer, F (2005) "A case study on the impact of Scrum on overtime and customer satisfaction". Agile Development Conference, IEEE Computer Society, p. 70-79.

[9] Nerur, S, Mahapatra, R, Mangalara, G (2005) "Challenges of Migrating to Agile Methodologies". Communication of the ACM, 48(5), pp.72-78.

[10] Svensson, H, Höst, M (2005) "Views from an organization on how agile development affects its collaboration with a software development team". Lecture Notes in Computer Science, v. 3547, Springer Verlag, Berlin, pp. 487–501.

[11] Sillitti, A, Ceschi, M, Russo, B, Succi, G (2005) "Managing uncertainty in requirements: a survey in documentation-driven and agile companies". In: Proceedings of the 11th International Software Metrics Symposium (METRICS).

[12] Abrahamsson, P, Conboy, K, Wang, X (2009) "'Lots Done, More To Do': the Current State of Agile Systems Development Research". European Journal of Information Systems, 18(4), pp. 281-284.

[13] Shahrbanoo, M, Ali, M, Mehran, M (2012) "An Approach for agile SOA development using agile principals". International Journal of Computer Science & Information Technology (IJCSIT), Vol 4, No 1, February, pp.237-244.

[14] Chow, T, Cao, DB (2008) "A survey study of critical success factors in agile software projects". The Journal of Systems and Software, v.81, pp. 961–971.

[15] Lee, G, Xia, W (2010) "Toward Agile: An Integrated Analysis Of Quantitative And Qualitative Field Data On Software Development Agility". MIS Quarterly, volume 34, N°1, pp. 87-114.

[16] Abbas, N., Gravell, A.M., Wills, G.B. (2010) "Using Factor Analysis to Generate Clusters of Agile Practices (A Guide for Agile Process Improvement)," Agile Conference (AGILE), vol., no., pp.11-20, 9-13 Aug. 2010. doi: 10.1109/AGILE.2010.15

[17] So, C. and W. Scholl (2009) "Perceptive Agile Measurement: New Instruments for Quantitative Studies in Pursuit of the Social-Psychological Effect of Agile Practices". Agile Processes in Software Engineering and Extreme Programming: 83-93.

[18] Santos, MA, Bermejo, PHS, Oliveira, MS, Tonelli, AO (2011) "Agile Practices: An Assessment of Perception of Value of Professionals on the Quality Criteria in Performance of Projects". Journal of Software Engineering and Applications, v. 04, p. 700-709.

[19] Sletholt, M, Hannay, J, Pfahl, D, Langtangen, H (2005) "What do we know about agile practices in scientific software development"? Computing in Science & Engineering, pp. (99):1.

[20] Asnawi, A.L, Gravell, A.M, Wills, G.B (2012) "Factor Analysis: Investigating Important Aspects for Agile Adoption in Malaysia" AGILE India (AGILE INDIA), vol., no., pp.60-63, 17-19, doi: 10.1109/AgileIndia.2012.13

[21] Puhl, S, Fahney, R (2011) "How to assign cost to 'avoidable requirements creep': A step towards the waterfall's agilization". In: Requirements Engineering Conference (RE), 19th IEEE International.

[22] Eckfeldt, B, Madden, R, Horowitz, J, Grotta, E (2005) "Selling Agile: Target-Cost Contract". Agile Development Conference (ADC '05), pp.160-166.

[23] Asnawi, A.L, Gravell, A.M, Wills, G.B (2012) "Emergence of Agile Methods: Perceptions from Software Practitioners in Malaysia" AGILE India (AGILE INDIA), vol., no., pp.30-39, 17-19, doi: 10.1109/AgileIndia.2012.14

[24] Lindvall, M, Muthig, D, Dagnino, A, Wallin, C, Stupperich, M, Kiefer, D, et al. (2004) "Agile software development in large organizations". Computer, 37 (12), 26–34.

[25] Jaakkola, H, Thalheim, B. (2010) "Framework for high-quality software design and development: a systematic approach". IET Software 4(2), 105-118.

[26] Jung, CF (2004) Metodologia Para Pesquisa & Desenvolvimento: Aplicada a Novas Tecnologias, Produtos e Processos. Axcel Books, Rio de Janeiro.

[27] Biffl, S, Aurum, A, Boehm, B, Erdogmus, H, Grünbacher, P (2006) Value-Based Software Engineering. Springer-Verlag, 388 p.

[28] Johnson, RA, Wichern, DW (2007) Applied multivariate statistical analysis. Prentice Hall, Edition 6th, 800 pp.

[29] Hair, Joseph F et al. (2009) Análise multivariada de dados. 6th ed. Bookman, Porto Alegre, 688 p.

[30] Mingoti, SA (2005) Análise de dados através de métodos de estatística multivariada: uma abordagem aplicada. Editora UFMG.

[31] Malhotra, NK (2001) Pesquisa de Marketing: uma orientação aplicada. Editora Bookman, Porto Alegre.

[32] Maroco, J (2007) Análise Estatística com a utilização do SPSS. 3rd ed. Silabo, Lisboa, 822 pp.

[33] Field, A. (2005). Discovering Statistics Using SPSS, Sage.

## AUTHORS

**Mariana de Azevedo Santos** is BSc in Information Systems at the Universidade Federal de Lavras (UFLA) in 2011 and systems analyst at Mitah Technologies, an IT start-up focused on high-tech solutions for processes and products traceability, with experience in development of ERP systems and mobile solutions. Also has experience as researcher in Software Engineering, specifically in the following fields: agile development, mobile application development, human-computer interaction and multivariate statistical data in software engineering.

**Paulo Henrique de Souza Bermejo** is an associate professor in the Department of Computer Science at the Universidade Federal de Lavras (UFLA). Dr. Bermejo graduated in the information system area, and has a doctorate degree in Engineering and Knowledge Management from the Federal University of Santa Catarina and is a post doctorate at Bentley University in Massachusetts (USA), where he has researched open innovation in software companies. He has served as an instructor in training courses and has consulted with institutions in the private and public sectors. He is certified in the IT governance framework COBIT. He has 15 years experience in the coordination and development of corporate and IT strategic projects and consulting for national and international institutions in the public and private sectors. He is co-author of the book *Risk Management In Software Projects: Based On Reference Process Models*, published by Ciência Moderna in Brazil. He has also published dozens of articles in scientific journals and conferences, and has worked on the construction of 24 software products. He worked at UFLA as Associate Dean for Graduate Studies from 2009 to 2011.

**Marcelo Silva de Oliveira** started his studies in Electrical Engineering at the Universidade Federal de Minas Gerais (UFMG) in 1979, but received his bachelor's degree in Agricultural Engineering at the Universidade Federal de Lavras (UFLA) in 1985. He earned a master's degree in Statistics at the Universidade Estadual de Campinas (UNICAMP) in 1991, and a doctorate degree of Engineering (Industrial Engineering) at the Universidade de São Paulo (USP) in 2000. Oliveira has more than 60 articles in scientific journals and conferences and a book on statistics fundamentals. He is currently an associate professor at the Universidõe Federal de Lavras (UFLA). He teaches Probability and Statistics in the undergraduate and graduate programs and his research interests include geostatistics, statistical quality control, and statistics fundamentals.

**Adriano Olímpio Tonelli** is currently a researcher in the Department of Computer Science at the Universidade Federal de Lavras (UFLA) and MSc candidate in Management at Department of Management and Economics at the Universidade Federal de Lavras (UFLA). Currently BSc in Computer Science and MBA in IT Governance at Universidade Federal de Lavras (UFLA), Tonelli is working on IT Governance, IT Service Management and Software Innovation. A teacher of the MBA Executive in IT Governance at the Universidade Federal de Lavras (UFLA), Tonelli has worked on consulting projects involving strategic planning, information security management, software development and IT governance in public and private organizations.

**Enio Júnior Seidel** has a Ph.D. degree in Statistics and Agricultural Experimentation at the Universidade Federal de Lavras (UFLA) and is assistant professor of Statistics at the Universidade Federal do Pampa (UNIPAMPA). He has a undergraduate degree in Mathematics from the Universidade Regional Integrada do Alto Uruguai e das Missões (URI) in 2004, a expertise course in Applied Statistics at the Universidade Federal de Santa Maria in 2005, another expertise course in Statistics and Quantitative Modelling at the Universidade Federal de Santa Maria in 2010 and a master degree in Industrial Engineering at Universidade Federal de Santa Maria in 2009.